# Mind the ground: A Power Spectral Density-based estimator for all-terrain rovers

Giulio Reina[1, 2], Antonio Leanza[2], Annalisa Milella[3], Arcangelo Messina[2]

[1]Department of Mechanics, Mathematics & Management, Polytechnic of Bari, Via Orabona 4, 70125, Bari, Italy

[2]Department of Engineering for Innovation, University of Salento, Via per Arnesano, 73100 Lecce, Italy

[3]Institute of Intelligent Industrial Technologies and Systems for Advanced Manufacturing, National Research Council, via G. Amendola 122/O, 70126 Bari, Italy.

Abstract

There is a growing interest in new sensing technologies and processing algorithms to increase the level of driving automation towards self-driving vehicles. The challenge for autonomy is especially difficult for the negotiation of uncharted scenarios, including natural terrain.

This paper proposes a method for terrain unevenness estimation that is based on the power spectral density (PSD) of the surface profile as measured by exteroceptive sensing, that is, by using a common onboard range sensor such as a stereoscopic camera. Using these components, the proposed estimator can evaluate terrain on-line during normal operations. PSD-based analysis provides insight not only on the magnitude of irregularities, but also on how these irregularities are distributed at various wavelengths. A feature vector can be defined to classify roughness that is proved a powerful statistical tool for the characterization of a given terrain fingerprint showing a limited sensitivity to vehicle tilt rotations. First, the theoretical foundations behind the PSD-based estimator are presented. Then, the system is validated in the field using an all-terrain rover that operates on various natural surfaces. It is shown its potential for automatic ground harshness estimation and, in general, for the development of driving assistance systems.

# 1. Introduction

Terrain profile plays a critical role in the dynamic response of off-road vehicles, influencing directly the traction, comfort, and handling performance. In extreme cases, it can endanger the vehicle possibly leading to entrapment or rollover. For example, one of the main causes for farm tractors overturns is the lack of assessment of operational risks due to irregular terrain [1]. Another notable example is the difficulty encountered by the NASA/JPL rovers during the exploration of Mars surface. In 2005, Opportunity became embedded in a dune of loosely packed drift material [2], whereas a similar embedding event experienced by Spirit in 2010 led to the end of its mobility operations [3].

Latest advances in sensing technologies have made available on the market new moderate-cost sensors with improved capabilities for 3D reconstruction of the surrounding environment that have paved the road to novel perception systems. They can assist the driver or improve the level of driving automation towards self-driving vehicles and autonomous robots. This aspect is especially important for off-road applications[1] where the environment is not structured, and the properties of the supporting surface are not available beforehand. In this work, an approach is presented for automatic roughness estimation of the terrain observed from a distance using an onboard range sensor. In the context of this research, a stereocamera is used as the available sensor modality. However, the proposed system is perception solution agnostic, that is, it requires as input a generic 3D point cloud that can be generated locally by alternative sensors including Lidars or depth cameras or that can be already available as a global map. The 3D reconstructed terrain surface is statistically analyzed via power spectral density (PSD) that provides a powerful tool to characterize the magnitude and distribution of its geometric properties.

The proposed method works on-line during normal maneuvering that is an advantage with respect to traditional approaches for surface estimation that have two main limitations: they are usually limited to paved roads and they are often performed through a time-consuming sampling and a successive off-line data processing. Therefore, the development of methods for automatic in-road/off-road terrain roughness classification represents an interesting challenge that would extend applications in different fields including service robotics and the automotive field, paving the way to new generations of vehicles that are sensitive to the surface unevenness [4]. There are several practical issues, which are encountered when applying the proposed approach in off-road scenarios. These issues are addressed and their effects on the terrain roughness characterization are discussed throughout the paper.

This approach may be integrated and used in conjunction with other modern advanced driving assistance systems (ADAS) to further increase the degree of driving automation towards fully autonomous vehicles.

The research is presented in the paper as follows: Section 2 presents a survey of the methods for terrain characterization available in the literature explaining the differences and the advantages of the proposed approach. Section 3 explains methods adopted in this research and the material used for testing and validation. Section 4 presents the theoretical foundation behind the proposed approach and introduces the concept of roughness parameters. In Section 5, experimental results obtained from the proposed method in real scenarios are presented and discussed. Section 6 draws final remarks and future improvements.

## 2. Literature review

This section surveys relevant research, which has been devoted to the problem of terrain roughness estimation for ground vehicles. Two main approaches have been proposed in the Literature. The first and most common is to assume a mathematically tractable form of a statistical nature to model artificial terrains that can be used as input to vehicle dynamic simulations. The second approach consists of a direct measurement of the supporting surface without any a priori assumption.

---

[1]https://robohub.org/self-driving-cars-for-country-roads/

Research on terrain modeling was developed for example by Li and Sandu who proposed methods based on polynomial chaos [5], and Autoregressive Moving Average (ARMA) [6]. There are also studies, which have modeled the terrain surface as fractals [7] and [8]. ISO standards [9] classify road type assuming the road-velocity profile as a purely random (i.e. Gaussian white noise) input. A similar model has been extended to irregular and deformable surfaces in [10].

Experimental estimation of the terrain surface can be performed via contact and contactless measurements. Examples of contact sensors are pin profilers and mesh-boards [11]. However, they are in disuse today since they tend to disturb or destroy the surface under investigation and provide biased and time-consuming estimates. Research that is more recent has been devoted to evaluate the extent of terrain irregularity by measuring the vibration behavior induced on the vehicle chassis via accelerometers attached to the body [12],[13] and [14], or to a trailer [15].

In contrast, modern surface digitizing methods employ non-contact sensors, including laser profilers [11], [16] and [17], and stereo imagers [18], [19] that have the advantage of measuring the terrain surface from a distance without actually traversing it. LIDAR scanners outperform stereo-cameras in terms of observation range and accuracy. However, they are expensive although new cheaper 3D LIDARs are now becoming available on the market. In addition, LIDARs may suffer from data sparseness that directly affects the quality of terrain surface reconstruction. They find widespread adoption for hazard detection and avoidance purposes [20]. In order to decrease costs, 2D Lidars can be used in combination with Global Positioning Systems (GPS) and Inertial Measurement Units (IMUs) blended in a Kalman filter for full 3D reconstruction of paved roads [21], [22], [23] and [24].

Stereo-cameras are more affordable and are becoming widely available on modern cars for advanced driving assistance system (ADAS) integration. They also ensure dense 3D reconstruction at short/medium range. In the context of this research, a stereo rig is used as the available sensor modality that represents a trade-off solution between cost and performance.

Traditionally the root mean squared elevation (RMSE) has been used as a means for terrain characterization [6] and [25]. More recently, an extended set of geometric metrics has been also used including the mean elevation, the average slope, and the goodness of planar fit [26]. Although these features are simple and computationally efficient, they are not fully descriptive of a given terrain profile since they fail to account for the different content at the various wavelengths of the profile under inspection. In this research, the use of the power spectral density (PSD) is proposed to provide a more in-depth description of the terrain distinctive traits. The PSD-based characterization takes into account the spatial frequency content of the terrain, capturing underlying trends that cannot be appreciated by simple geometric features [27].

The application of such a technique for terrain unevenness characterization by a rough-terrain vehicle in the context of all-terrain driving represents an original contribution of this work.

## 3. Materials and Methods

### 3.1. System Architecture

For the extensive testing during its development, the proposed system was integrated with a customised all-terrain rover that is shown in Figure 1. The vehicle is 0.7 m long, 0.5 m wide, 1.2 m high with a 20 kg payload. Its sensor suite includes electrical current and voltage sensors to measure wheel mechanical torque, encoders to estimate wheel angular velocity, and an inertial measurement unit (IMU) (XSENS MTi-300) that outputs the vehicle angular rate and acceleration measurements, and estimates navigation body attitude (i.e., Euler angles: roll, pitch, yaw) through a firmware-implemented extended Kalman filter. A stereo camera (Point Grey XB3) mounted on a dedicated aluminium frame provides three-dimensional reconstruction with RGB colour data of the surrounding environment.

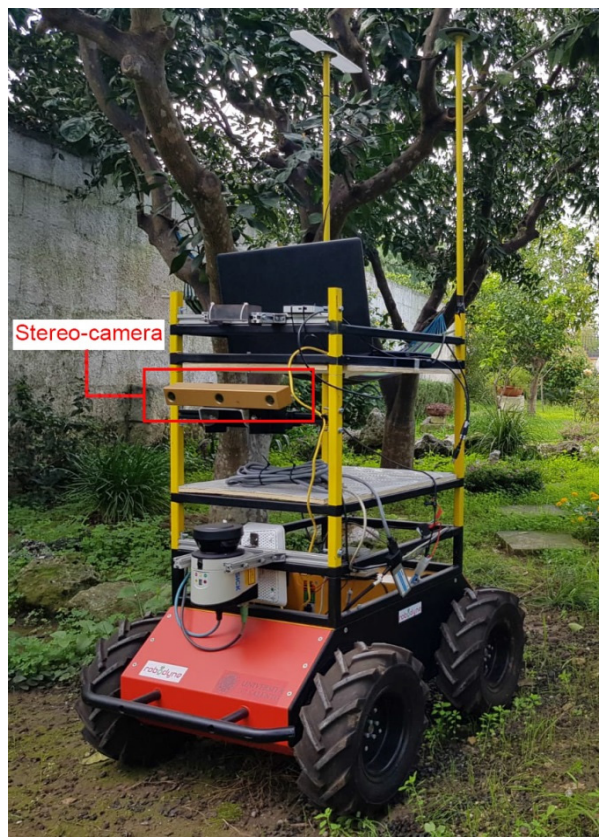

Figure 1: The all-terrain robot used during the system development and testing

## 3.2. Acquisition setup and methodology

Figure 2 shows a block diagram explaining the rationale behind the proposed approach for terrain estimation. As the vehicle travels on an unknown surface, the onboard stereo camera generates a three-dimensional reconstruction of the environment in front of the vehicle. Attention is given to the rolling terrain patch 1-m ahead of the robot with dimensions matching approximately the vehicle chassis (i.e., 0.90 x 0.90 m). In this region of interest, the average accuracy of the 3D stereo reconstruction results in 0.004 m with a 0.002 m standard deviation. The sampling spatial frequency, $B$, is 0.008 m.

In our implementation, a 2 Hz refresh rate of the stereo device (maximum frame rate of 16 fps) was adopted ensuring a smooth update of the terrain patch at the operating travel speed of 0.5 m/s., i.e., two successive patches have approximately 70% overlap. It should be noted that the travel velocity was kept at an approximately constant value to eliminate, on first approximation, the influence of velocity on the three-dimensional reconstruction of the terrain patch. The acquisition set-up was chosen to better fit the robot characteristics (e.g., maximum travel speed of 0.7 m/s) and it may be easily adapted to different vehicles and higher travel velocities. For example, for robots having velocities in the order of 5 m/s, the maximum sampling rate of the stereocamera would be necessary (i.e., 16 fps). It is also worth mentioning that the technology development that have followed the recent interest in driverless cars have made available on the market new ultrafast cameras that record up to 120 frames per second[2].

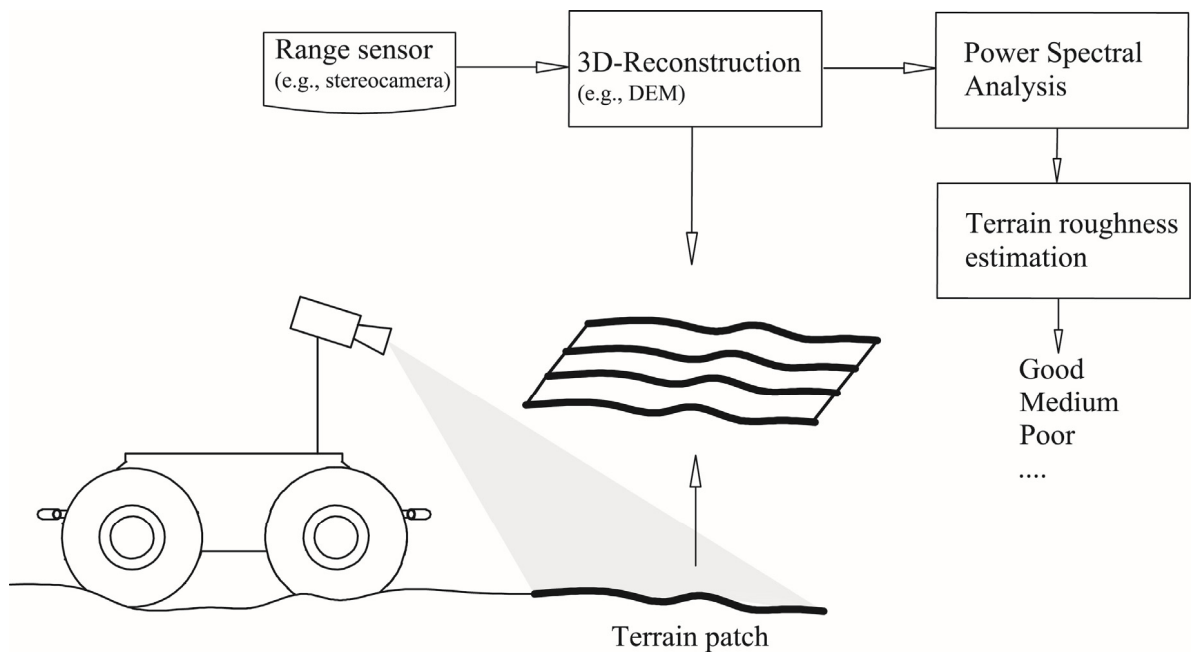

Figure 2: Pipeline of the proposed system for terrain estimation using the PSD analysis of the terrain profile

The proposed algorithm operates on a 3D point cloud acquired by the robot while in motion. In the context of this research, three-dimensional reconstruction of the environment is obtained using a stereo processing algorithm that was developed in previous research by the authors. Here, the main steps of the algorithms are briefly recalled, whereas the interested reader is referred to [27] for more details.

---

[2]https://www.greencarcongress.com/2017/02/20170217-ntu.html

> *Disparity map computation*: to compute the disparity map, the Semi-Global Matching (SGM) algorithm is used. This algorithm combines concepts of local and global stereo methods for accurate, pixel-wise matching with low runtime.
>
> *3D point cloud generation*: as the stereo pair was calibrated both intrinsically and extrinsically, disparity values can be converted to depth values and 3D coordinates can be computed in the reference camera frame for all matched points. A statistical moving window filter is applied to reduce noise and discard outliers. In addition, frequency components with wavelength longer than the scan length are removed by applying a regression analysis with one-order polynomials through a de-trending manipulation [9].
>
> *Patch of interest extraction*: Only the 3D points that fall into the region of attention are retained and used for successive statistical analysis.

It is worth to note that the 3D point cloud obtained following the previous steps is referred to the vehicle reference frame. During negotiation of irregular terrain, the vehicle continuously changes its attitude. Therefore, in order to express the point cloud in a fixed inertial reference system, a transformation is required to compensate for the corresponding vehicle tilt angles. In principle, the knowledge of the current set of Euler angles (pitch-roll-yaw) would be necessary via, for example, an IMU. However, the proposed system based on PSD analysis does not entail the use of additional onboard sensors, being little affected by the vehicle tilt movements (refer to Section 5.3), which is one of its strengths.

## 3.3. Test field

The system was tested in a rural environment where three main surfaces were present:

- ploughed terrain: vineyard terrain broken and turned over
- compact terrain: unbroken agricultural land. It is a compact and relatively hard terrain that can be typically found in olive groves
- gravel: unconsolidated mixture of white/gray rock fragments or pebbles.

Sample images of the different terrain types are shown in the first column of Figure 3 with overlaid the projection of the boundaries of the current patch of interest. It can be seen that ploughed terrain is distinguishable from all other terrain classes due to the presence of furrows and ridges that are expected to result in distinctive geometric characteristics with respect to compact terrain and, partially to gravel. The second column of Figure 3 shows the corresponding 3D patch of interest as obtained by the stereo reconstruction.

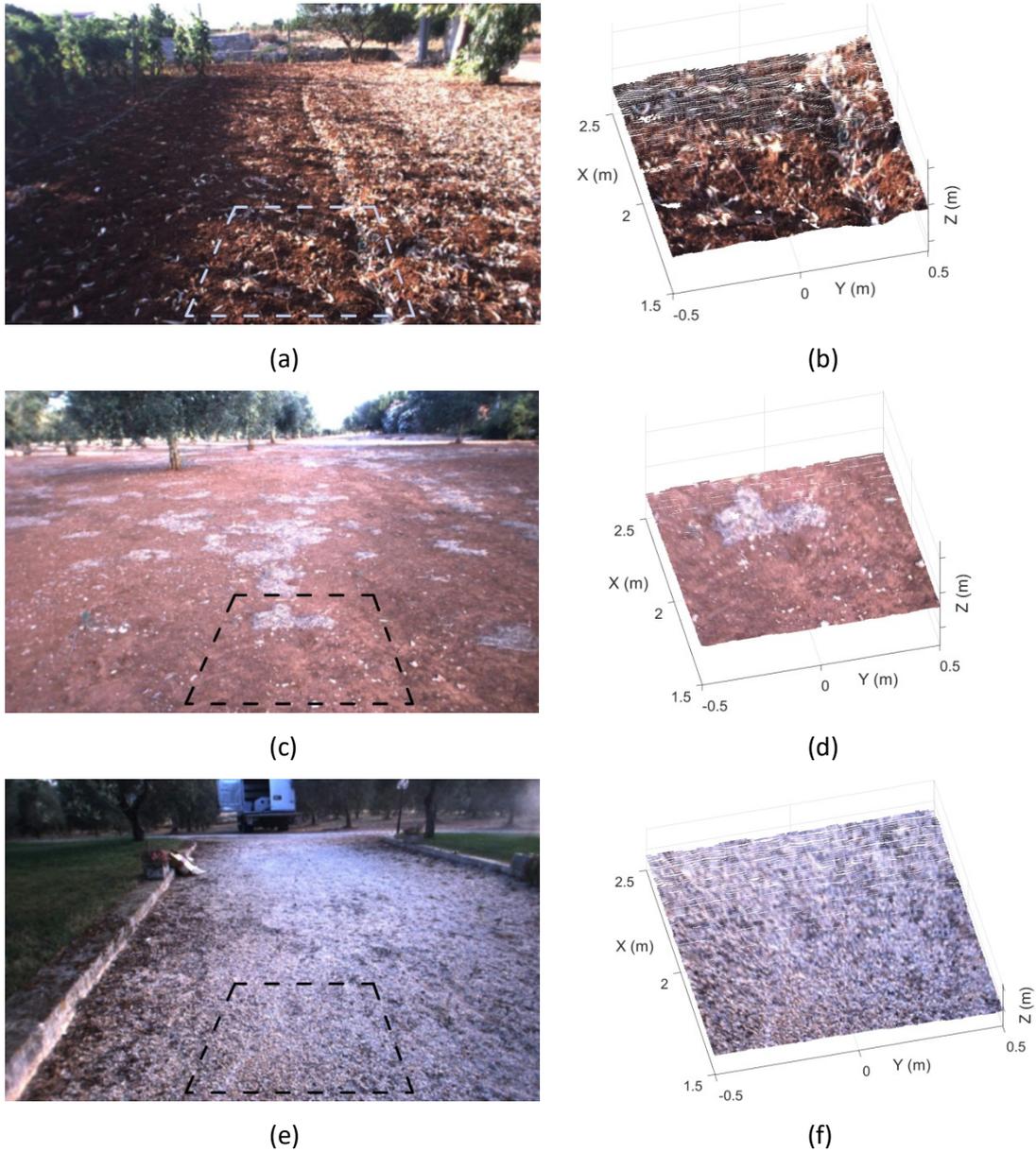

Figure 3: Terrain types: (a)-(b) ploughed agricultural terrain; (c)-(d) well-packed agricultural terrain; (e)-(f) gravel. Left: original sample images with overlaid the boundaries of the inspection window. Right: corresponding 3D terrain patches with RGB content obtained as output from the stereo vision algorithm.

## 4. Terrain characterization

Assume that terrain elevation can be considered to be a realization of a random process that unfolds as a function of the traveled distance. Then, terrain unevenness, which is responsible for the continuous dynamic response of a ground vehicle, can be described in statistical terms, i.e. in terms of power spectral density. The PSD can be formally defined as the Fourier transform of the autocorrelation function $A(\xi)$ of the elevation profile:

$$\phi(\Omega) = \frac{1}{2\pi}\int_{-\infty}^{+\infty} A(\xi)e^{-i\Omega\xi}d\xi \qquad (1)$$

with $\xi$ the spatial shift and $\Omega$ the wavenumber. For more mathematical details the reader is referred to the broad Literature in this field, for example to [29].

The first practical challenge to face is to define the appropriate finite waveband of interest, $\Omega_b$. The lower wavelength limit is set as twice the value of the spatial sampling frequency (or step grid), $B$, in compliance with the Nyquist's theorem [30]. The high wavelength limit is fixed by the scan range that is by the length of the search window that in this implementation is set $L = 1\,m$ approximately equal to the vehicle wheelbase. Therefore, recalling that the waveband $\Omega_b$ is defined as $2\pi$ times the reciprocal of the profile wavelengths, measured in $\frac{rad}{m}$:

$$\forall k \in \left[1, \frac{n}{2}\right] : \Omega_k = \frac{2\pi k}{nB} \tag{2}$$

with *n* number of points of each longitudinal profile and *B* the spatial sampling frequency. The search window length is $L = nB$, so the lower and upper bound of $\Omega_b$ will be respectively:

$$\Omega_1 = \frac{2\pi}{nB} = \frac{2\pi}{L} \quad and \quad \Omega_l = \frac{\pi}{B} \tag{3}$$

being the longest and shortest profile wavelength, respectively, $\lambda_{max} = L$ and $\lambda_{min} = 2B$. Thus, $\Omega_b$ contains a number $l$ of elements, with $l = \frac{n}{2}$.

The Bartlett-Welch method [31] is used to calculate the one-sided PSD of a given elevation profile, $\phi$, over the waveband $\Omega_b$. As an example, Figure 4 shows the PSD (solid grey line) of a single longitudinal profile extracted from a patch acquired during the traverse of packed agricultural terrain. As seen from this figure, the PSD typically decreases with increasing wavenumber (spatial frequency). It is common to approximate the spectral density curve of the ground profile by an exponential equation, which gives a straight-line fit on a log-log plot of the PSD spatial frequency diagram (solid black line in Figure 4):

$$\hat{\phi}(\Omega) = R\Omega^w \tag{4}$$

where $w$ and $R$ are constants that relate, respectively, to the slope and scaling of the fit line, $\hat{\phi}$. $w$ is referred to as *waviness,* and its magnitude indicates how quickly the energy of the surface decreases as the wavenumber increases. $R$ is related to the overall energy of the surface and its value is estimated at the reference wavenumber $\Omega_0 = 1\,\frac{rad}{m}$, i.e. $R = \hat{\phi}(\Omega_0)$. Waviness and overall energy are chosen as characteristic properties of a certain elevation profile and collectively referred to as roughness parameters in the remainder of this work.

The second practical issue is to extend the concept of roughness parameters from a single profile to the entire patch. Figure 5 shows the scatter plot of the roughness parameters as obtained from all elevation profiles included in the given patch. Note that the overall energy $R$ is reported in log scale to better highlight the inversely proportional correlation with the waviness.

The analytical expressions of all linear regression fits of the patch profiles can be conveniently expressed in a matrix compact form as

$$ln(\widehat{\boldsymbol{\phi}}(\boldsymbol{\Omega})) = ln(\boldsymbol{R}) \cdot \boldsymbol{1}^T + (\boldsymbol{w} \cdot \boldsymbol{1}^T) \cdot \ln(\boldsymbol{\Omega}) \tag{5}$$

where $\boldsymbol{R}$ and $\boldsymbol{w}$ are $m \times 1$ matrixes of the roughness parameters pertaining to the patch composed by $m$ profiles and $\boldsymbol{1}$ is ones vector of dimension $l \times 1$.

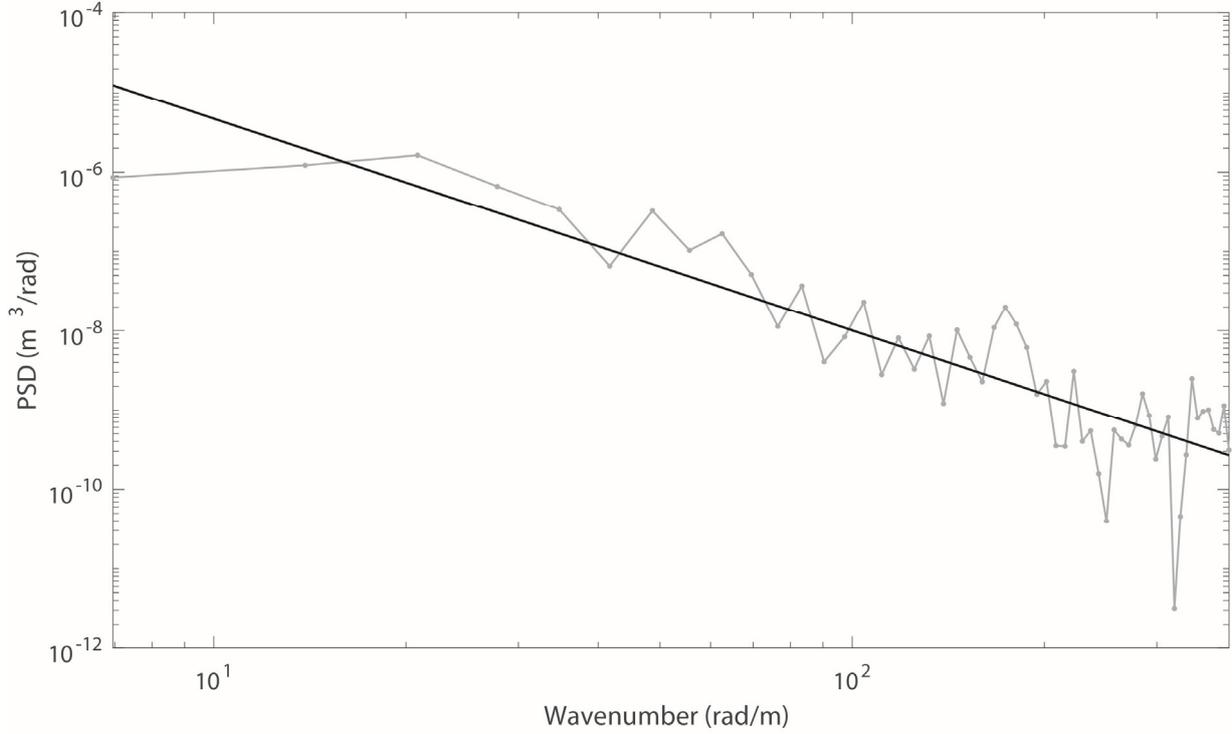

Figure 4: PSD of a single elevation profile extracted from a patch of packed terrain along with its linear regression fit

By placing $\boldsymbol{b} = ln(\boldsymbol{R})$ and applying the linearity property of the expected value to Eq. (5):

$$E[ln(\widehat{\boldsymbol{\phi}}(\boldsymbol{\Omega}))] = E[\boldsymbol{b}] + E[\boldsymbol{w}] \cdot E[ln(\boldsymbol{\Omega})] \tag{6}$$

and passing from the logarithmic form to the explicit one:

$$\widehat{\boldsymbol{\phi}}(\boldsymbol{\Omega}) = e^{E[\boldsymbol{b}]} \boldsymbol{\Omega}^{E[\boldsymbol{w}]} \tag{7}$$

one gets the roughness parameters representative of the whole patch, $(e^{E[\boldsymbol{b}]}, E[\boldsymbol{w}])$ that are marked by a black circle marker in Figure 5. The feature vector defines statistically the geometric fingerprint of the window under inspection and it used to classify a certain terrain in this research.

Since $e^b$ is a *non – linear* function of the random variable $\boldsymbol{b}$, attention should be paid for its expected value and variance evaluation; it can be proved (refer to the Appendix) that, for any *non–linear* function $Y = g(X)$ of random variable $X$:

$$\begin{cases} \hat{\mu}_Y = E[Y] = g(E[X]) \\ \hat{\sigma}_Y^2 = \left[\dfrac{\partial}{\partial X} g(X)_{|\mu_X}\right]^2 \sigma_X^2 \end{cases} \tag{8}$$

The first row of Eq. (8) just confirms the expression of the expected value for $\boldsymbol{R}$ in Eq. (7), whereas the second row is useful to estimate the uncertainty (standard deviation) of $\boldsymbol{R}$.

As a final remark, it should be noted that the PSD inherently makes the assumption that the profile is statistically stationary, which is not the case for natural terrains. Thus, there could be variations in the PSD estimation for shorter or longer profile lengths and the spatial frequency range of the meaurment shall be always reported.

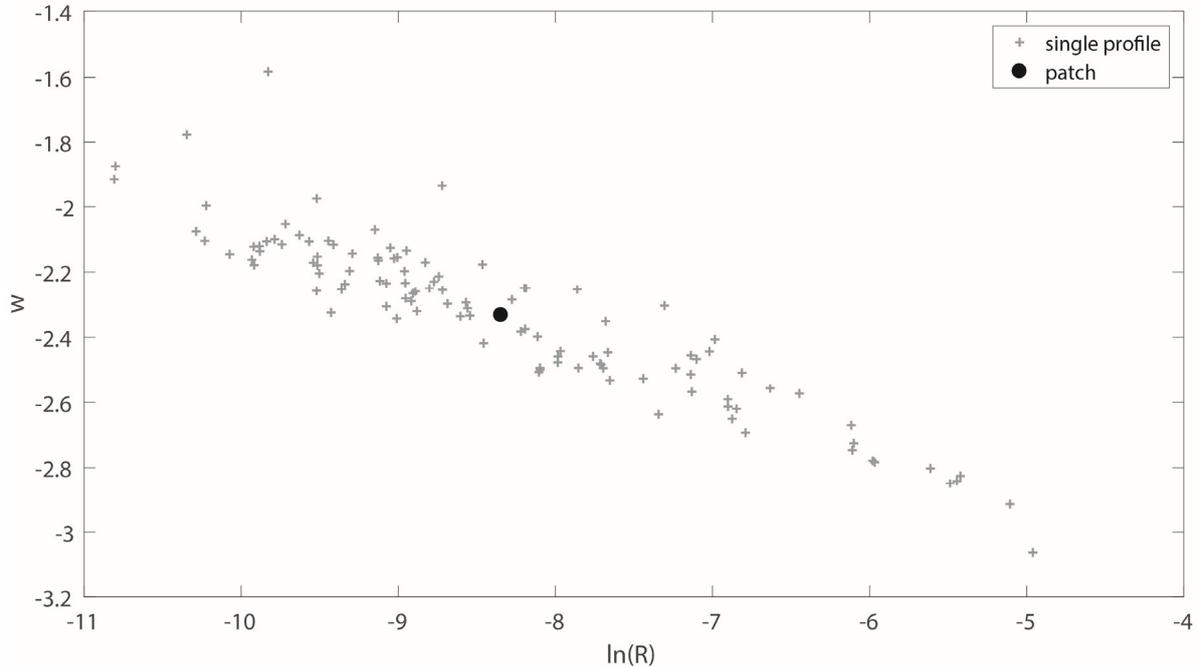

Figure 5: Scatter plot of the terrain roughness parameters obtained from the patch elevation profiles. The black circle marker indicates the average value representative of the whole patch.

## 5. Results

### 5.1 Field evaluation

This section presents results as obtained from the proposed system integrated with the rover shown in Figure 1. First, a set of experiments was performed on stretches of approximately uniform terrain, namely *ploughed* and *compact agricultural terrain*. Then, the rover was driven along a course comprising sections of different surface types.

Figure 6 shows a sample image acquired during the traverse of *ploughed* terrain along with the corresponding 3D reconstruction of the inspection window. The rover was remotely controlled to follow approximately a straight line of about 42 m, driving parallel to a vineyard edge. Estimation of roughness parameters as obtained from the system along the path is shown in Figure 7. The upper plot shows the estimated overall energy marked by a solid black line along with the associated uncertainty, expressed as standard deviation and denoted by a grey shaded area. For convenience, the reference values for paved roads as indicated by the *ISO 8608* [8], are indicated as well, where ISO type A refers to very good asphalt, whereas ISO type F points to very poor asphalt. In the lower plot of Figure 7, the estimated waviness and its uncertainty are shown, respectively, by a solid black line and a grey shaded area. In order to get a quantitative evaluation of the unevenness extent for the investigated surface, the roughness parameters along the path were estimated, resulting in an average overall energy of $\bar{R} = 1734 \cdot 10^{-6} \frac{m^3}{rad}$ with a standard deviation of $1200 \cdot 10^{-6} \frac{m^3}{rad}$ and an average waviness $\bar{w} = -2.59$ with a standard deviation of 0.08.

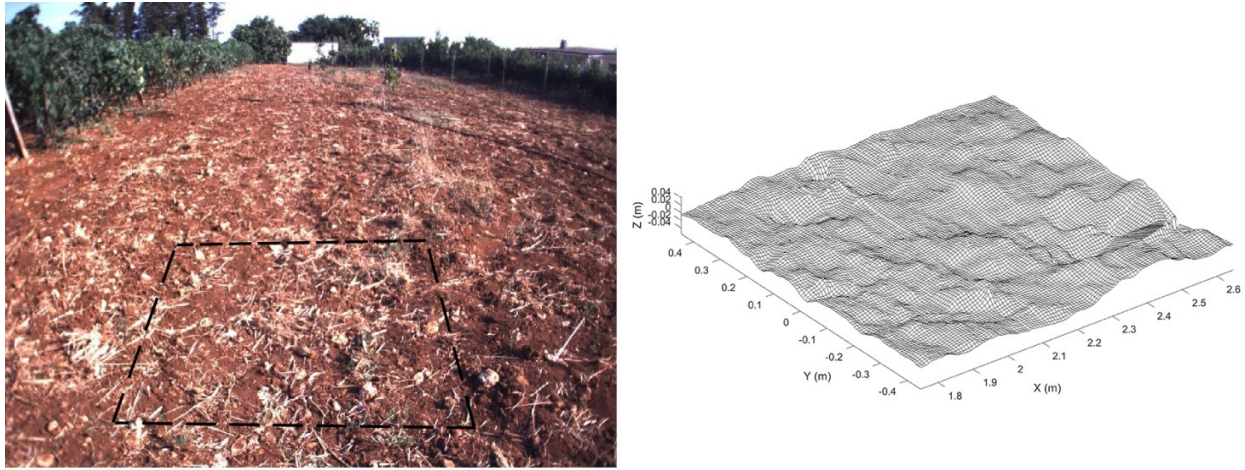

Figure 6: Sample image and 3D reconstruction of the corresponding inspection window: *ploughed agricultural* terrain

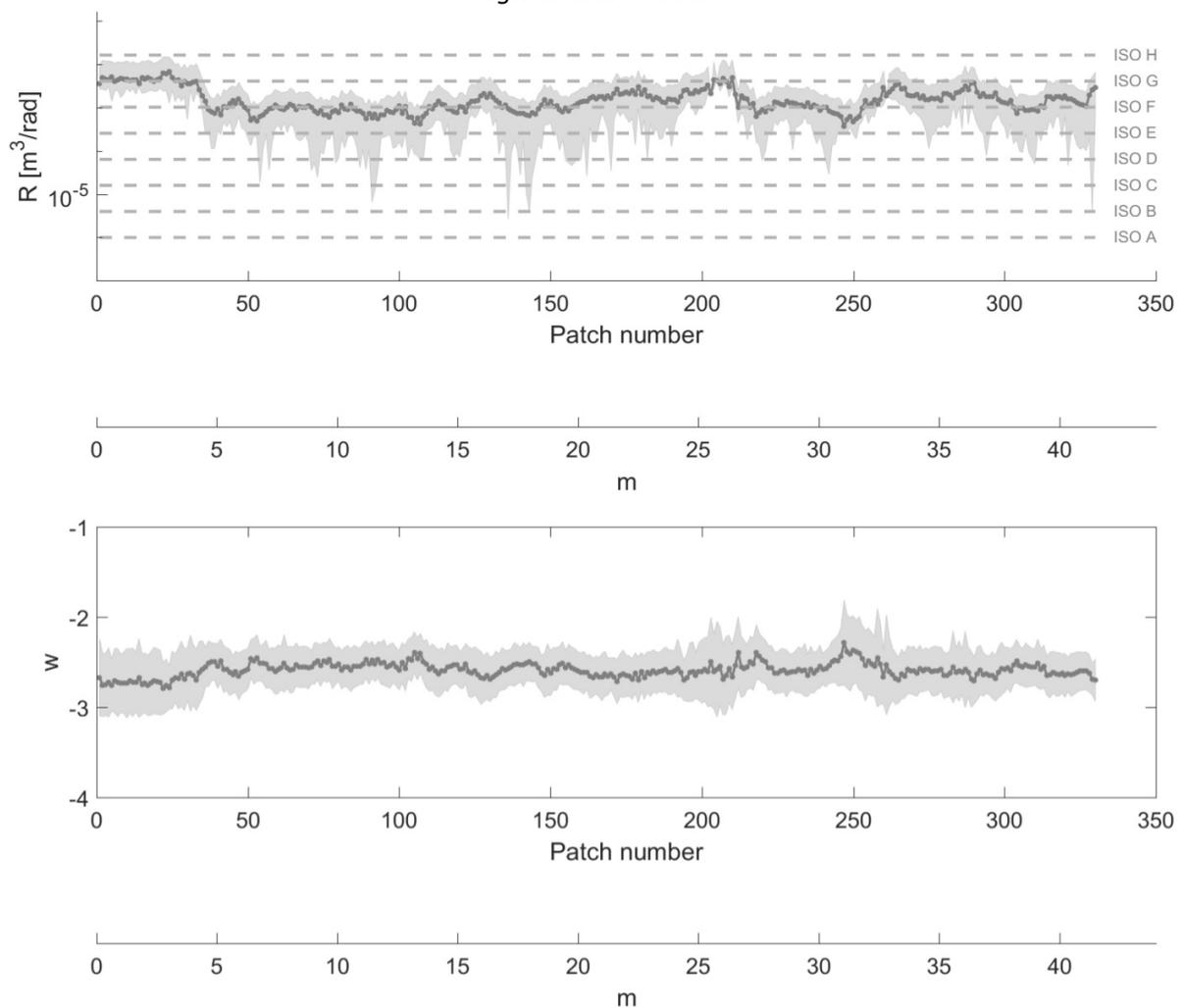

Figure 7: Roughness parameters $(R, w)$ as obtained from the system along a straight course on *ploughed agricultural* terrain

Another experiment was performed by driving the rover in an olive grove along an approximately straight path of 22 m on compact agricultural terrain where a lower degree of terrain roughness is expected. Figure 8 shows a sample image and the corresponding window of interest as reconstructed by the stereovision algorithm. Estimation of roughness parameters along with their statistical spread are shown in Figure 9. Again, estimation of the overall energy (waviness) is reported in the upper (lower) plot of Figure 9.

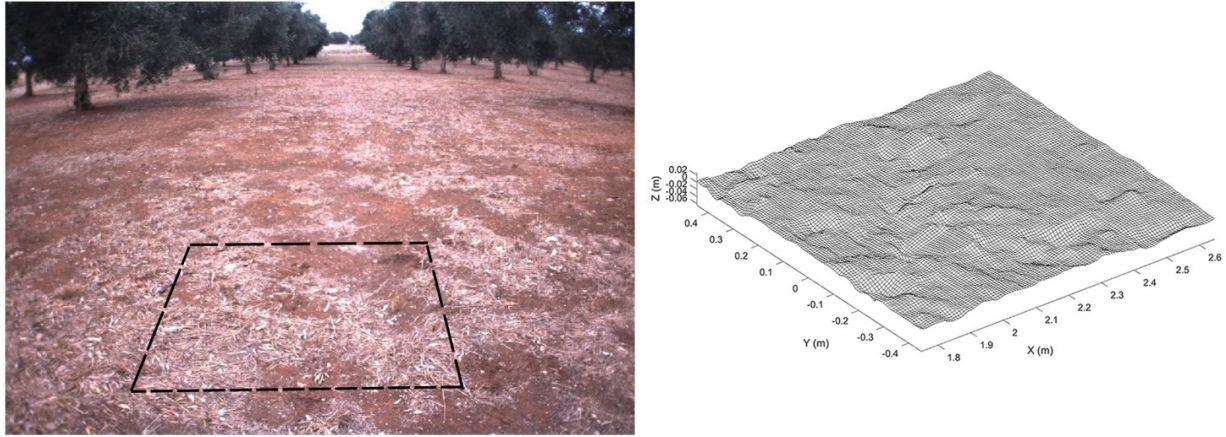

Figure 8: Sample image and 3D reconstruction of the corresponding inspection window: *compact* terrain

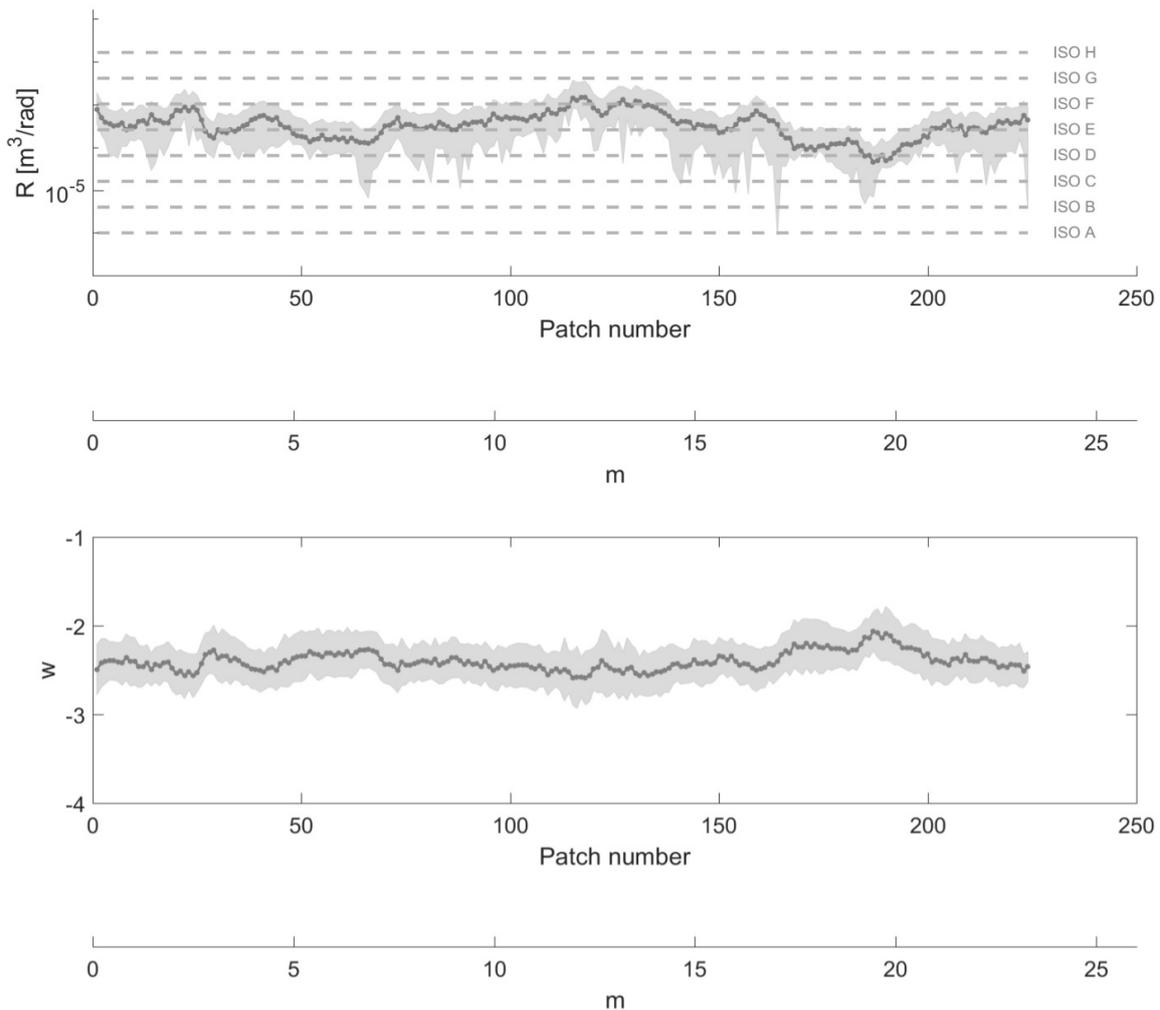

Figure 9: Roughness parameters $(R, w)$ as obtained from the system along a straight course on *compact agricultural* terrain

As seen from this figure, the estimated unevenness degree is significantly lower than ploughed terrain resulting in an average value of the roughness parameters over the whole path, respectively, $\bar{R} = 393 \cdot 10^{-6} \frac{m^3}{rad}$ with standard deviation $279 \cdot 10^{-6} \frac{m^3}{rad}$, and $\bar{w} = -2.40$ with standard deviation 0.10.

It is interesting to note that natural terrains show a larger energy decrease (i.e., waviness lower than -2) for increasing wave number than paved surfaces (i.e., waviness equal to -2). For a side-by-side comparison, the straight-line fits of the PSD frequency diagram for the two investigated terrains are plotted on a log-log graph and contrasted with the ISO road profiles, as shown in Figure 10. Natural terrains are characterized by higher energy at large wavelengths (low wavenumbers) compared to artificial man-made surfaces.

It also worth to say that the system performed consistently over various runs repeated on the same portion of terrain or different stretches of the same surface.

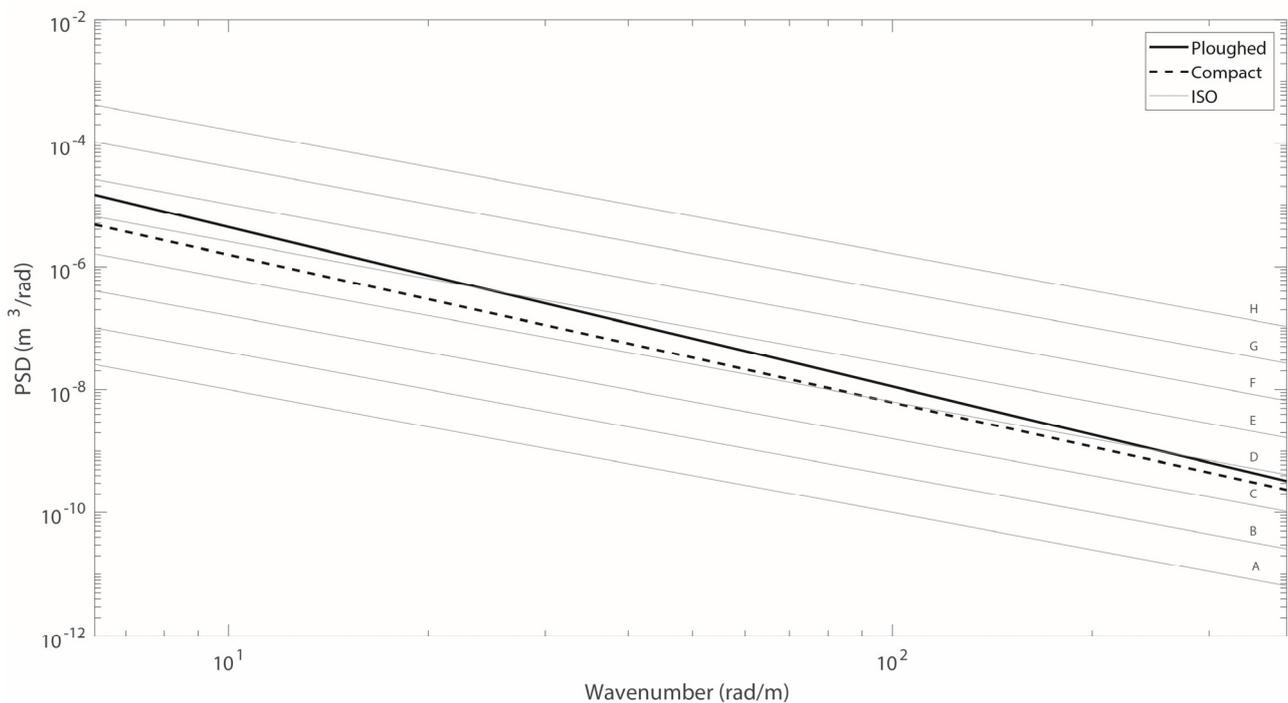

Figure 10: PSD curves for natural and paved ISO surfaces.

A second set of experiments was performed in order to evaluate the system ability to detect surfaces with different unevenness properties along the same course, as described in Figure 11. The rover started from a smooth concrete surface, then it moved on a dirt country road and ended on a gravel-like surface. It is interesting to look at the estimated roughness parameters that are collected in Figure 12. The rover encountered surfaces with increasing irregularity. The overall energy increased from ISO type C for the hard and smooth concrete pathway, to ISO type D on the well-beaten dirt road, and finally to ISO type E for the gravel-like surface. Similarly, waviness matched the typical value of ISO roads (i.e., approximately w=-2) when the rover was on the man-made concrete pathway, whereas it tended to decrease moving towards the dirt road (w≈-2.2) and the gravel (w≈-2.4).

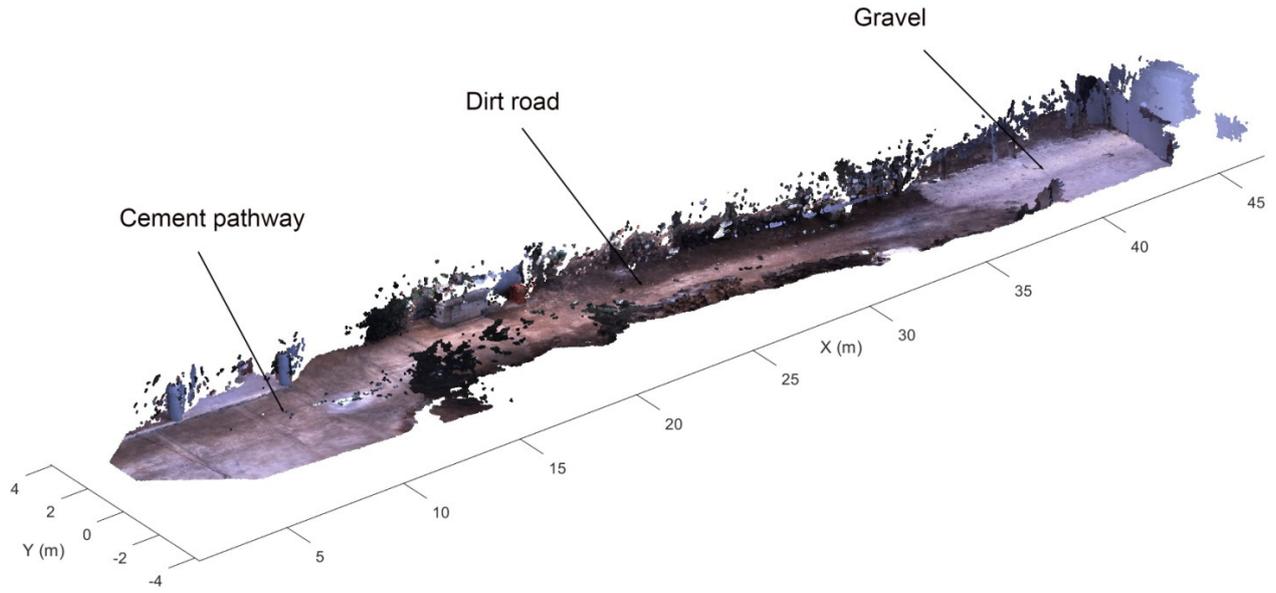

Figure 11: 3D reconstruction as obtained from the stereovision algorithm in a field trial along a course with mixed surfaces (*concrete – dirt road – gravel-like surface*)

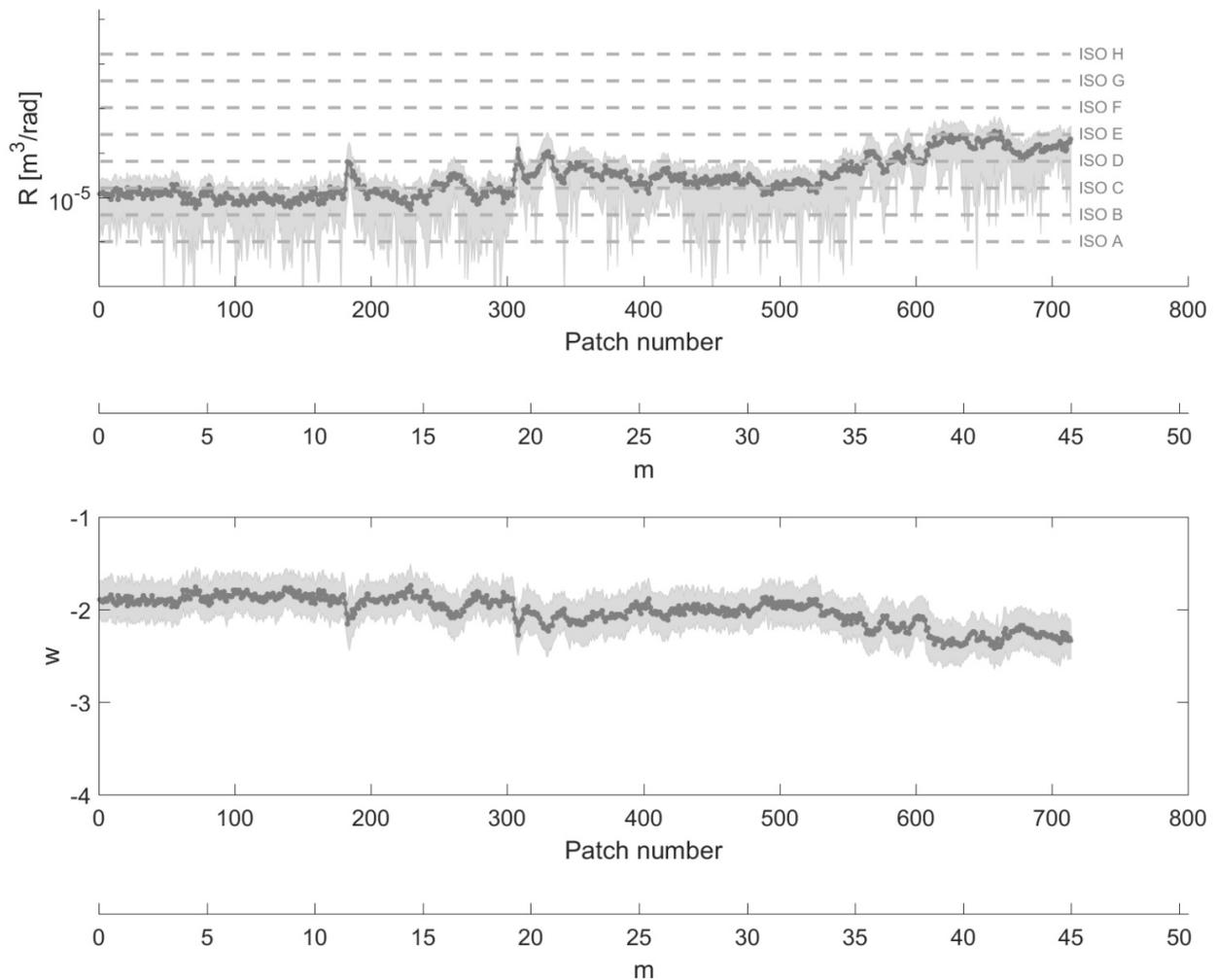

Figure 12: Roughness parameters $(R, w)$ as obtained from the system along a course with mixed terrains (*concrete – dirt road – gravel-like surface*)

Although, the primary interest of this research is the characterization of random properties of the terrain surface, isolated "defects" such as pits, protrusions, or ridges can also be detected. As an example, at about 12 m along the path on the concrete pathway, the system flags the presence of a manhole that is signalled by a spike in the roughness parameters, as highlighted in Figure 13 where the corresponding 3D reconstruction provided by the stereovision system is shown.

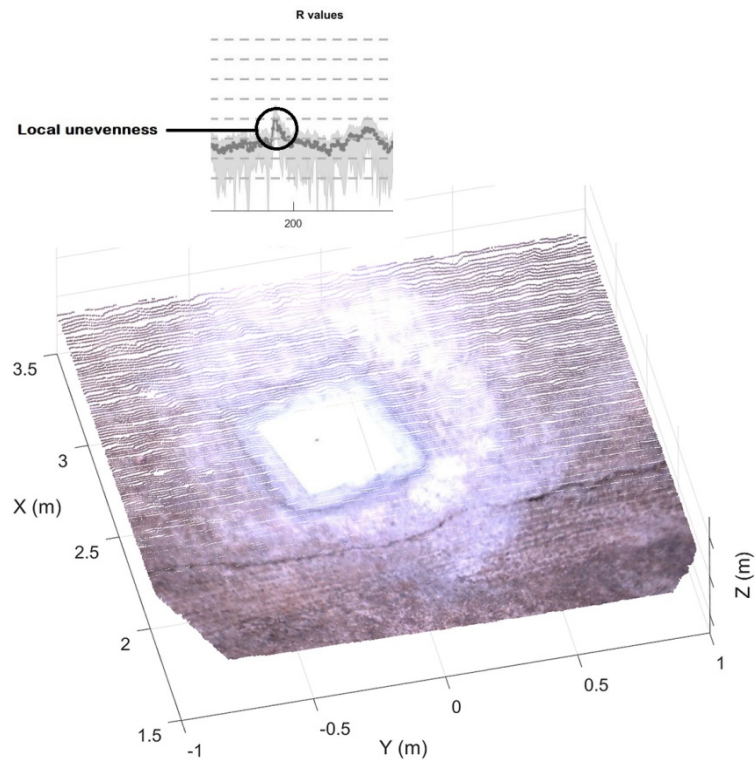

Figure 13: Detection of isolated "defects" such as a manhole along the concrete pathway.

Another experiment was performed in the same environment but driving the rover in a reverse order starting from the gravel-like surface and moving on the dirt road and finally to the concrete pathway. Figure 14 shows the results in terms of estimated roughness parameters as obtained by the system. As seen from this figure, the system consistently detected the changes from one surface to the other.

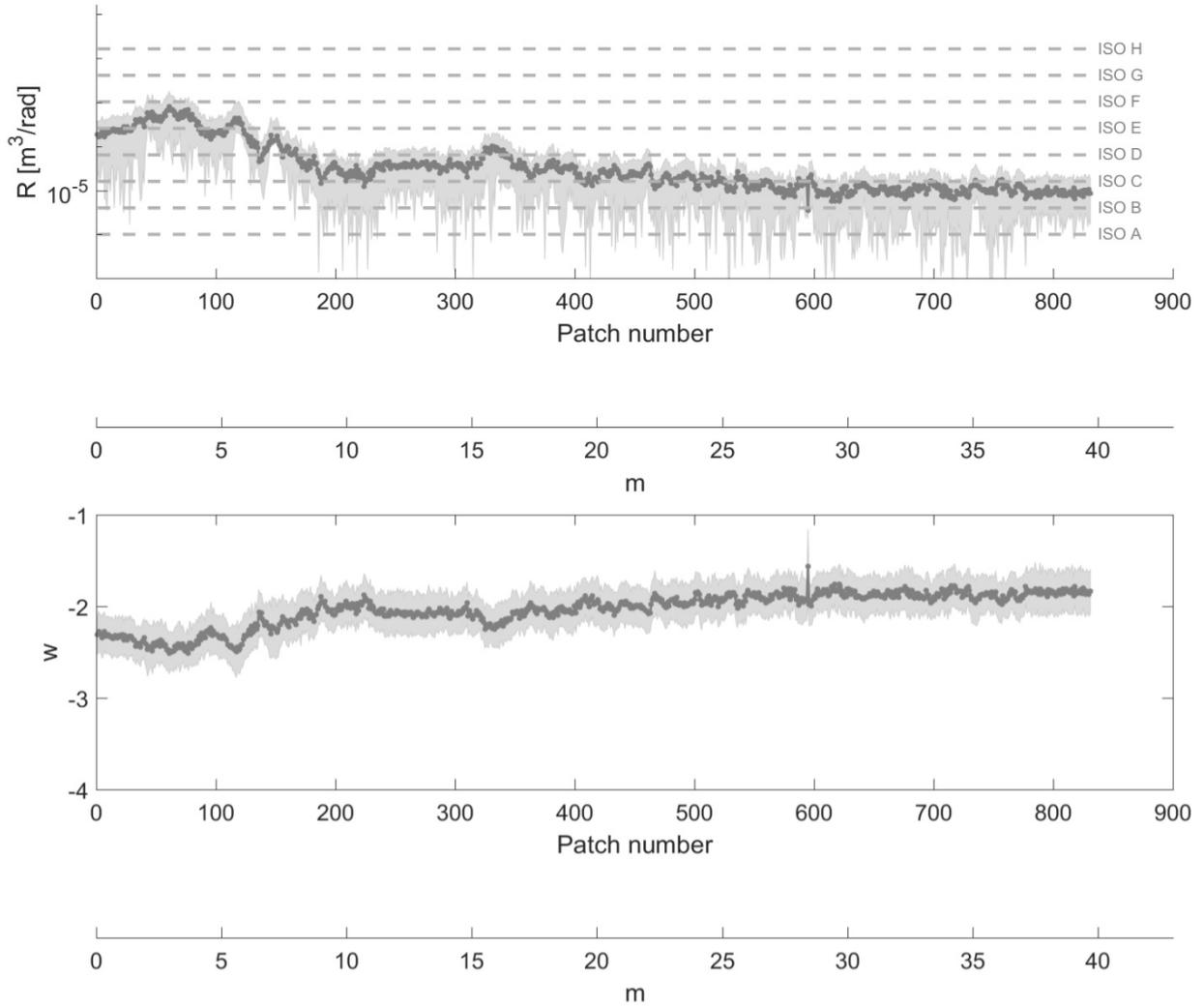

Figure 14: Roughness parameters $(R, w)$ as obtained from the system along a course with mixed terrains (*gravel-like surface– dirt road-concrete*)

## 5.2 System sensitivity

The accuracy of the surface topography is directly related with the uncertainty of the algorithm for environment reconstruction, as described in Section 3.2. It is especially important to understand the lowest degree of terrain roughness that the system can appreciate. In general, it is known that the relationship between root-mean-square (*rms*) and PSD of an elevation profile can be expressed as [32]

$$rms^2 = \int_{\Omega_1}^{\Omega_l} \phi(\Omega) d\Omega \qquad (9)$$

Drawing on *ISO 8608*, the road velocity profile is assumed as a Gaussian white random process, i.e., the PSD is equal to $\phi(\Omega) = \phi(\Omega_0) \left(\frac{\Omega}{\Omega_0}\right)^{-2}$. Note that in this case, the waviness is considered constant and equal to -2, and the roughness parameters reduce to the overall energy *R*. By solving the integral of Eq. (8), it is possible to link *R* directly with the elevation profile *rms*:

$$\phi(\Omega_0)_{sensor} = \frac{rms^2}{\frac{1}{\Omega_1} - \frac{1}{\Omega_l}} \cdot \frac{1}{\Omega_0^2} \tag{10}$$

For convenience, we can refer to the overall energy data reported in the *ISO 8608* for paved roads, in the waveband of interest, i.e. between $\Omega_1$ and $\Omega_l$. Corresponding values of *rms* obtained using Eq. (9) for the ISO profiles are collected in the second column of Table 1.

| ISO | $\phi(\Omega_0)\ 10^{-6}\ [m^3/\text{rad}]$ | rms [mm] |
|---|---|---|
| A | 1 | 0.37 |
| B | 4 | 0.75 |
| C | 16 | 1.50 |
| D | 64 | 2.99 |
| E | 256 | 5.99 |
| F | 1024 | 11.98 |
| G | 4096 | 23.95 |
| H | 16384 | 47.90 |

Table 1: RMS of the ISO 8608 road profiles in the waveband of interest

A set of experiments was performed by driving the rover on the flooring of an outdoor patio that we consider as our reference flat. Sample results obtained in this experiment are showed in Figure 15. The average *rms* obtained over various scans resulted in 2.22 mm. As seen from the second column of Table 1 with reference to the ISO profiles, the system sensitivity falls between ISO C and D. Therefore, surfaces with a lower extent of roughness cannot be differentiated by the system in its current implementation. Nonetheless, natural terrains for which the system is intended to, mostly fall in the region above type D, attesting to the feasibility of this approach.

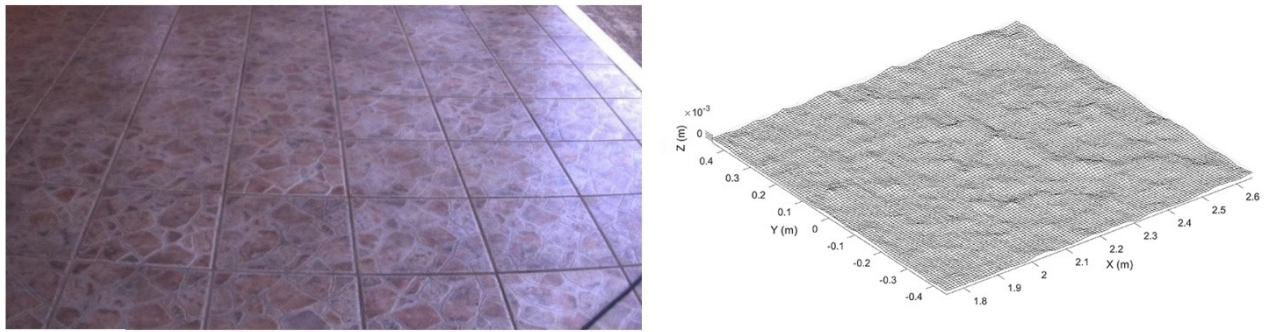

Figure 15: Sample image and 3D reconstruction of the corresponding inspection window: flooring of an outdoor patio

## 5.3. Influence of vehicle tilt

During the traverse of natural terrain, the rover is subject roll and pitch motions. In this section, the impact of tilt rotations is experimentally evaluated. The rover was controlled to follow an approximately 45 m-long straight path on agricultural ploughed terrain where the vehicle is expected to encounter significant tilting.

Figure 16 shows the roll, θ, and pitch, φ, angle as measured by the onboard IMU. The *rms* of the roll angle along the path resulted in 1.26° with a maximum value of 3.1°, whereas the *rms* of the pitch angle was 1.62° with a maximum of 4.46°.

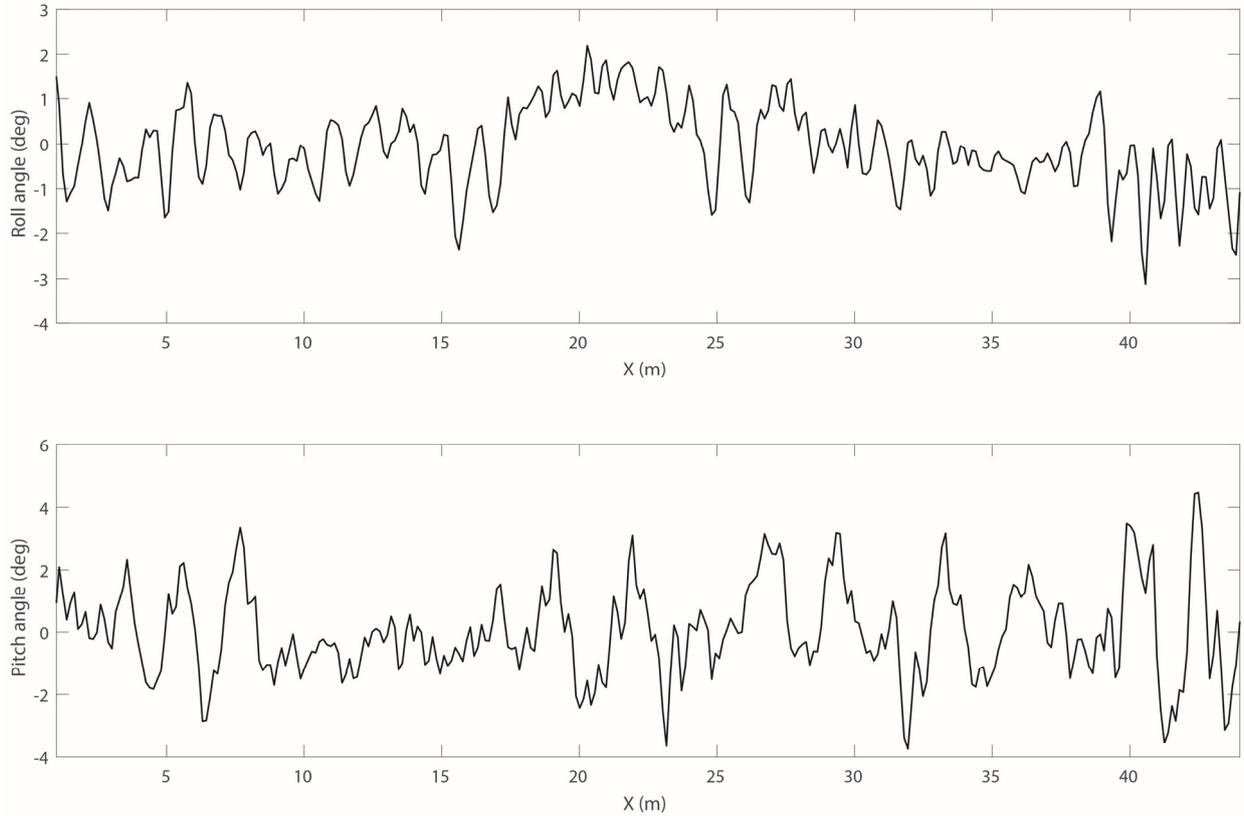

Figure 16: Roll and Pitch angles measured by the onboard IMU for ploughed agricultural terrain

Two different embodiments of the system for terrain roughness estimation are considered. In the first implementation, 3D point clouds generated by the stereovision algorithm are compensated for the vehicle tilt and transformed into the inertial fixed reference frame by applying the following rotation matrix:

$$R_v^w = \begin{bmatrix} \cos\varphi & \sin\varphi\sin\theta & \sin\varphi\cos\theta \\ 0 & \cos\theta & -\sin\theta \\ -\sin\varphi & \cos\varphi\sin\theta & \cos\varphi\cos\theta \end{bmatrix} \quad (11)$$

The second implementation does not correct the vehicle tilt and it operates directly on the 3D point cloud expressed in the vehicle reference frame.

Figure 17 shows the estimation of the roughness parameters as obtained from the two approaches that are denoted using a solid black line for the "compensated" approach, and a solid grey line for the "uncompensated" one. As apparent from this figure, the two implementations lead to similar result. The average relative error resulted respectively in 2.07% for the overall energy and 1.01% for the waviness. In the worst-case scenario, an error of 7.62% and 6.51% was registered, respectively for *R* and *w*. These results indicate that the estimation of the terrain roughness parameters was little affected attesting to the limited influence of vehicle tilting on the system performance. This aspect points to an advantage of the proposed approach that does not require additional sensors for vehicle tilt measurement and compensation. More investigation is required to evaluate the impact of extreme vehicle tilting due for example to higher travel velocities. However, this is left to future research.

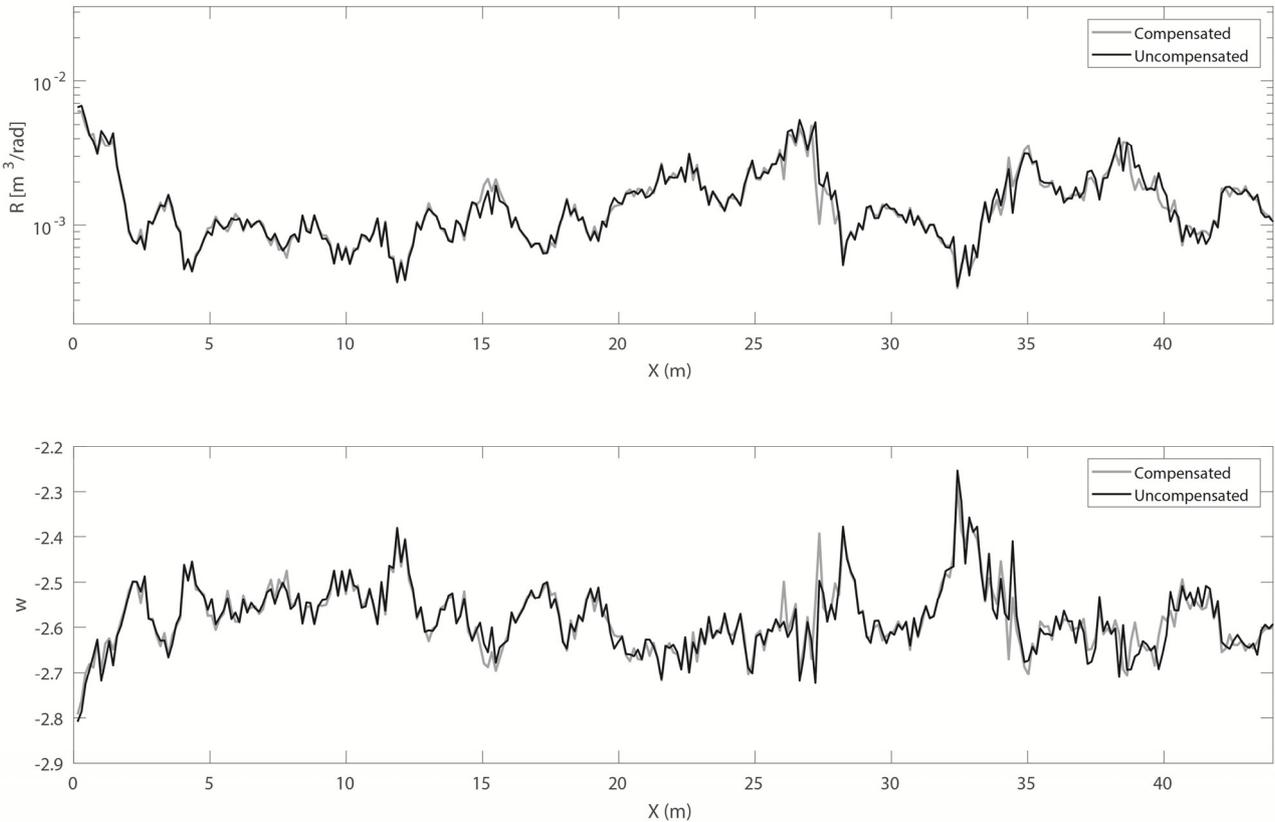

Figure 17: Estimation of roughness parameters $(R, w)$ as obtained from the system along a course on ploughed terrain using tilt compensation or raw stereo data.

## 5.4. Influence of robot speed

This Section investigates the impact on the system of the robot travel speed. A data set was acquired in the field by driving the vehicle with different constant velocities (e.g., 0.3-0.5-0.7 m/s) along a 50 m long straight path on ploughed terrain. In each of the three runs, terrain roughness parameters were estimated by the system keeping all other parameters (e.g., sampling rate and patch length) constant. Results are collected in Figure 18 showing that the distribution of the observations in the different tests is very similar. Therefore, the influence of the robot velocity on the system performance is rather limited. Obviously, other tests with significantly higher velocity would be necessary for a definitive conclusion.

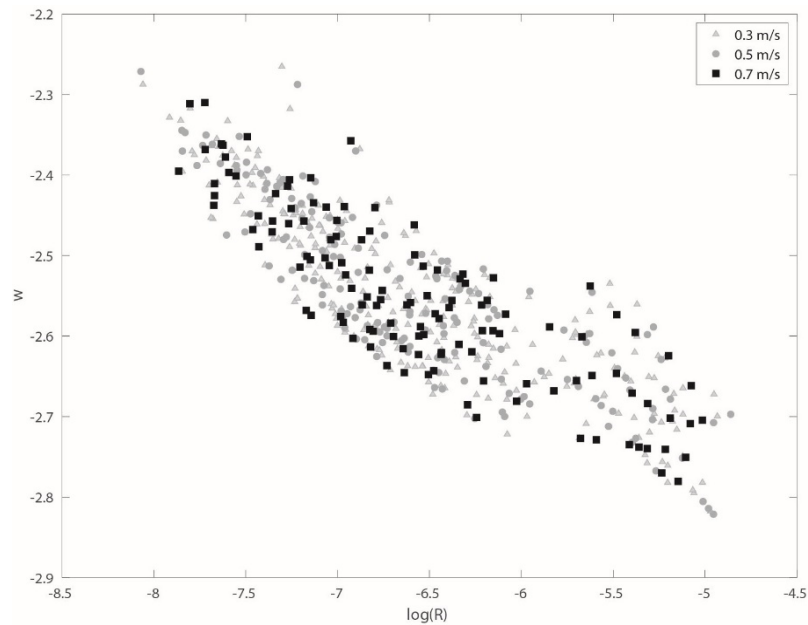

Figure 18: Estimation of roughness parameters $(R, w)$ as obtained from the system driving the robot with three different velocities along a 50 m long straight on ploughed terrain.

## 5.5. Terrain semantic labelling

One of the possible applications of the PSD-based estimator for terrain roughness is the development of driving assistance systems. A higher level of abstraction can be achieved in the displaying of terrain maps by introducing a new map layer based on the extent of surface roughness measured in front of the vehicle. Figure 19 collects the roughness parameters, as estimated by the system on various surfaces. The feature distribution indicates that the roughness parameters increase in magnitude proportionally to the extent of terrain unevenness. This suggests a semantic labelling where the successive search windows encountered by the rover are marked according to a colour map scaled on the estimated values of the roughness parameters, as indicated in the upper part of Figure 19. Three discrete levels of terrain roughness are considered: low, medium, and high.

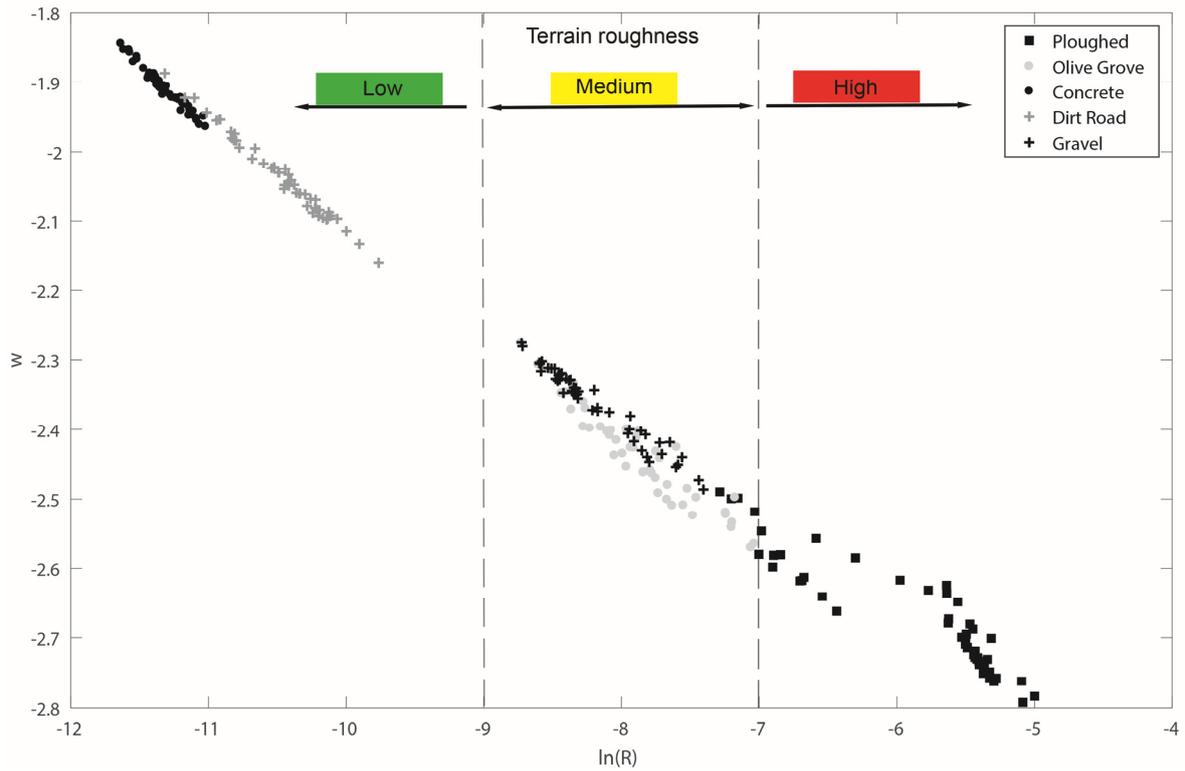

Figure 19: Roughness parameters $(R, w)$ as obtained from the system on various surfaces.

Then, results can be presented to the user with different informative layers as shown in Figure 20 for the first "mixed" experiment described in Section 5.1. The upper plot reports the 3D stereo-generated layer of the surveyed environment with the indication of the path followed by the rover denoted with a solid blue line, whereas the bottom plot shows the corresponding terrain roughness layer. Another example is shown in Figure 21 that refers to the traverse of ploughed agricultural terrain, instead.

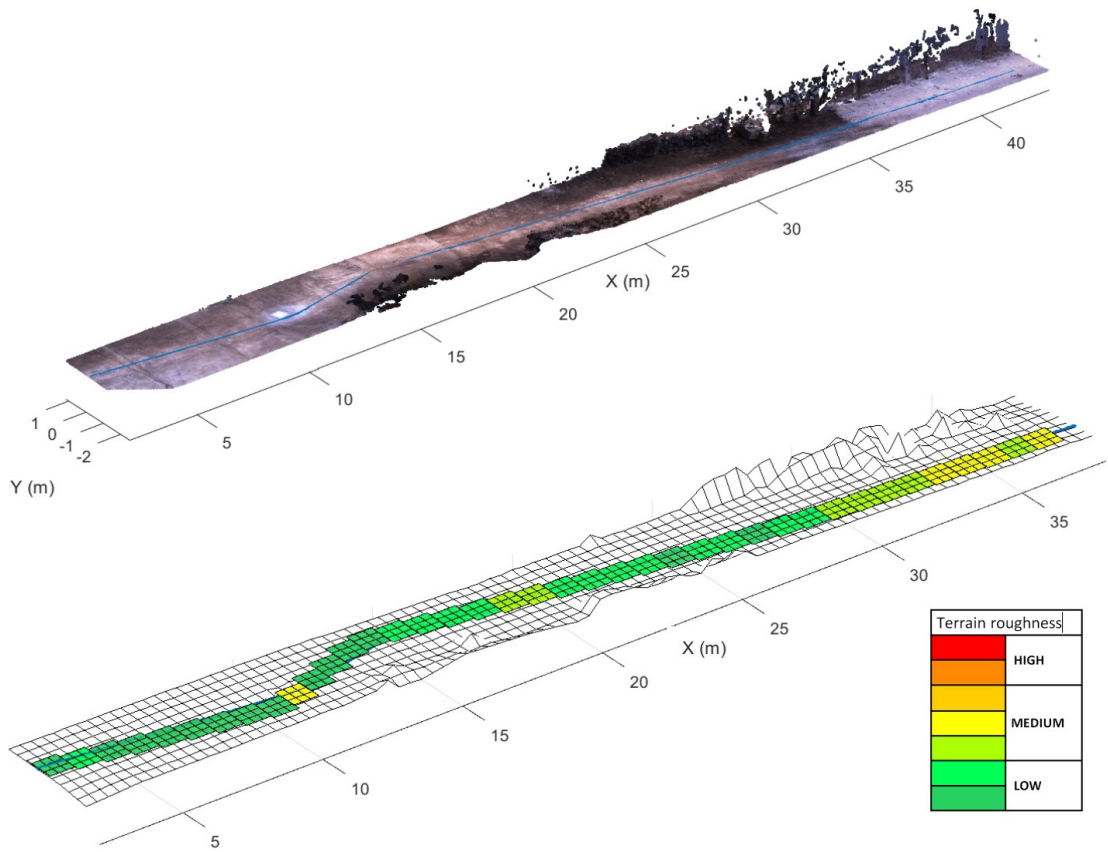

Figure 20: Display of the results obtained from the system using two informative layers: 3D stereo-reconstruction of the environment (upper plot), terrain roughness map (lower plot): mixed course.

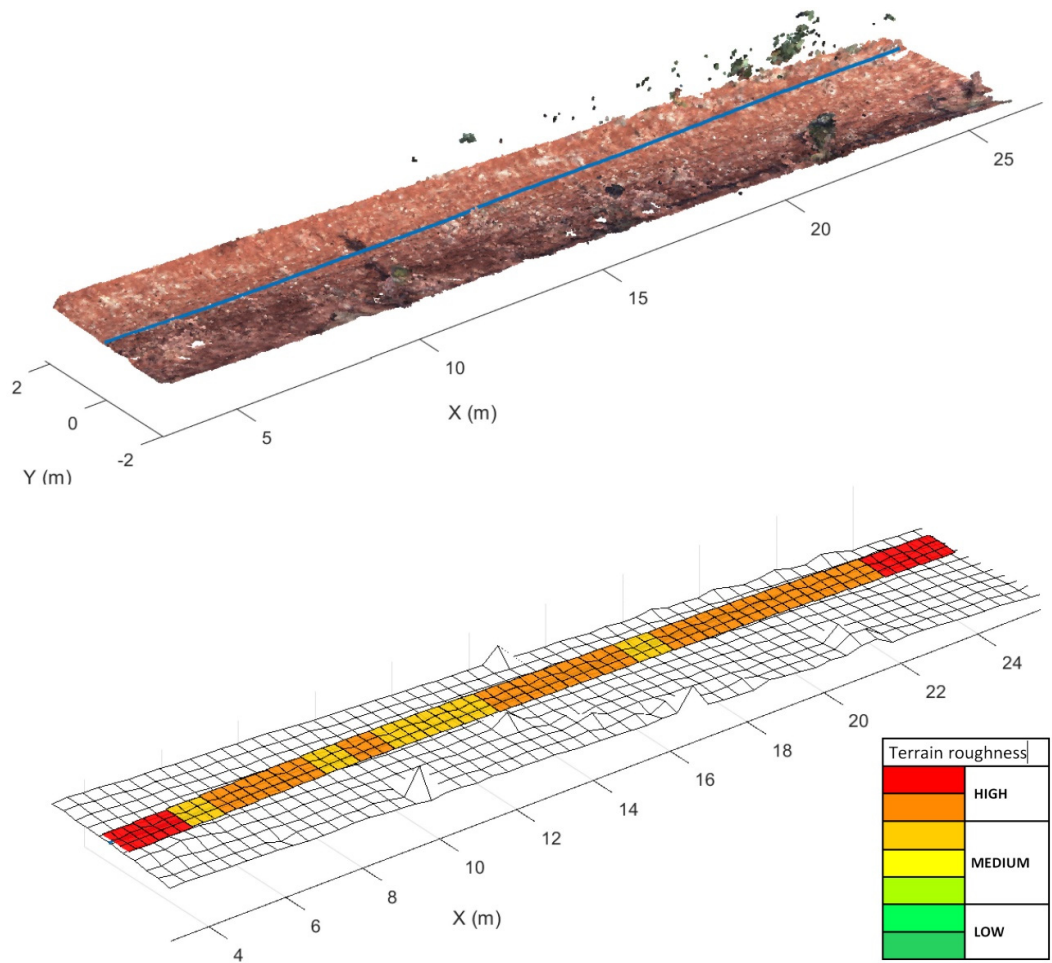

Figure 21: Display of the results obtained from the system using two informative layers: 3D stereo-reconstruction of the environment (upper plot), terrain roughness map (lower plot): ploughed terrain.

# 6. Conclusion and future developments

This paper presented a method to automatically estimate the roughness extent of the ground surface intended for intelligent vehicles with a high degree of autonomy. It is based on the analysis of the power spectral density of the terrain ahead as reconstructed by a stereocamera, but any range imaging sensor can be alternatively employed. The PSD-based analysis suggested the use of two parameters, overall energy and waviness, collectively referred to as roughness parameters, to infer the characteristic fingerprint of a given surface. The paper presented the theoretical foundation of the proposed approach along with a discussion about the practical issues for its implementation. It was validated in the field using an all-terrain rover operating on various uniform and mixed surfaces, including ploughed and packed agricultural terrain, gravel, and dirt road. In all experiments, the system performed consistently showing its ability to detect the extent of random irregularity of a given terrain and to differentiate between various surfaces. The system performance was evaluated in terms of sensitivity and the influence to vehicle tilting was proved to be limited with errors always less than 8%.

In addition, the distribution of roughness parameters over various investigated terrains indicated a new high-level semantic labeling approach to display the results for driving assistance purposes.

As a final remark, it should be noted that measurements obtained from various surfaces or different robots can be compared only when they refer to the same inspection waveband. Therefore, it is always important to claim the waveband of interest for each observation. The accuracy in the estimation of the roughness parameters is tightly connected with the quality of the surface reconstruction in front of the robot. Therefore, any uncertainty in the reconstructed 3D point cloud will reflect poorly on the measurement. For stereovision systems, special attention should be devoted to the calibration stage and to errors introduced by adverse lighting conditions or shadowing. Another limitation of the proposed system is that it is intended to provide estimates of "geometric" terrain irregularity that can be sensed remotely using range sensors. No information can be gathered about "non-geometric hazards" including loosely packed soil or soft soil covered by leaves. Future developments will focus on the analysis of the system performance under extreme roll and pitch motions. Stereodata will be also combined with other sensor sources, for example accelerometers to measure the vibration response of the vehicle as induced by the terrain irregularity. This could help the system to filter out the influence of superficial vegetation or leaves that can bias the estimation of the surface irregularity. A near to far learning system could be also developed where the terrain properties estimated in the near range are predicted in the far range using RGB color and stereo-range data.

## Acknowledgements

The financial support of the projects: Autonomous DEcision making in very long traverses (ADE), H2020 (Grant No. 821988), and Agricultural inTeroperabiLity and Analysis System (ATLAS), H2020 (Grant No. 857125), is gratefully acknowledged.

## References

[1] G. Reina, A. Milella, R. Rouvere, M. Nielsen, R. Worst, and M.R. Blas, Ambient Awareness for Agricultural Robotic Vehicles, Biosystems Engineering 146 (2016) 114-132.

[2] R. Cowen, Opportunity Rolls out of Purgatory, Science News 167(26) (2005) 413.

[3] L. Grossman, Atom & cosmos: Spirit stuck, but in good spot, Science News 177(5) (2010) 7.

[4] P. Papadakis, Terrain traversability analysis methods for unmanned ground vehicles: A survey, Engineering Applications of Artificial Intelligence (2013), 26(4) 1373-1385.

[5] L. Li, and C. Sandu, Modeling of 1-D and 2-D Terrain Profiles using a Polynomial Chaos Approach, In: 16th International Conference of International Society for Terrain-Vehicle Systems (ISTVS), (2008) Turin, Italy.

[6] L. Li, and C. Sandu, Modeling and Simulation of 2D ARMA Terrain Models for Vehicle Dynamics Applications, Journal of Commercial Vehicles 116(7) (2007).

[7] J.G. Howe, D.C. Lee, L.P. Chrstos, et al., Analysis of Potential Road/Terrain Characterization Rating Metrics, In: SAE Commercial Vehicle Engineering Congress and Exhibition, (2004) Rosemount, IL.

[8] Q. Wang, J. Guo and Y. Chen, Fractal Modeling of Off-Road Terrain Oriented to Vehicle Virtual Test, Journal of Zhejiang University SCIENCE A 7 (2006) 287-292.

[9] ISO 8608 (1995), Mechanical vibration -road surface profiles- reporting of measured data.

[10] G. Reina, A. Leanza and A. Messina, On the vibration analysis of off-road vehicles: Influence of terrain deformation and irregularity, Journal of Vibration and Control 24(22) (2018) 5418-5436.

[11] F. Mattia, M.W.J. Davidson, T. Le Toan et al., A comparison between soil roughness statistics used in surface scattering models derived from mechanical and laser profilers, IEEE Transactions on Geoscience and Remote Sensing 41 (2003) 1659–1671.

[12] P. Komma, C. Weiss and A. Zell, Adaptive Bayesian Filtering for Vibration -based Terrain Classification, IEEE International Conference on Robotics and Automation, (2009) Kobe, Japan

[13] C. Weiss, H. Fröhlich and M. Stark, Vibration-based Terrain Classification Using Support Vector Machines, International Conference on Intelligent Robots and Systems, October 9 – 15 (2006) Beijing, China.

[14] C.A.Brooks and K. Iagnemma, Vibration-Based Terrain Classification for Planetary Exploration Rovers, IEEE Transactions on Robotics 21(6) (2005).

[15] H. Imine, Y. Delanne and N. Msirdi, Road profile input estimation in vehicle dynamics simulation, Vehicle System Dynamics 44 (4) (2006) 285-303.

[16] L. Zhixiong, C. Nan, U.D. Perdok et al., Characterization of Soil Profile Roughness, Biosystems Engineering 91 (2005) 369–377.

[17] F. Darboux and C.H. Huang, An instantaneous-profile laser scanner to measure soil surface microtopography, Soil Science Society of America Journal 67 (2003) 92–99

[18] A. Chilian and H. Hirschműller, Stereo Camera Based Navigation of Mobile Robots on Rough Terrain, IEEE/RSJ International Conference on, (2009) Louis, USA

[19] M. Bellone, G. Reina, A. Messina, A new approach for terrain analysis in mobile robot applications, IEEE International Conference on Mechatronics, (2013), 225-230.

[20] S. Thrun, M. Montemerlo, D. Dahlkamp et al., (*Intelligent Robots and Systems* 2006). Stanley: The robot that Won the DARPA Grand Challenge, Journal of Field Robotics 23(9) (2006) 661-692

[21] C. Urmson, C. Ragusa, D. and Ray, A Robust Approach to High-Speed Navigation for Unrehearsed Desert Terrain, Journal of Field Robotics 23(8) (2006) 467-508.

[22] J.V. Kern and J.B. Ferris, Development of a 3-D Vehicle-Terrain Measurement System Part I: Equipment Setup, In: Joint North America, Asia-Pacific ISTVS Conference and Annual Meeting of Japanese Society for Terramechanics, (2007) Fairbanks, AL.

[23] S.M. Wagner, J.V. Kern, W.B. Israel et al., Development of a 3-D Vehicle- Terrain Measurement System Part II: Signal Processing and Validation, In: Joint North America, Asia-Pacific ISTVS Conference and Annual Meeting of Japanese Society for Terramechanics, (2007) Fairbanks, AL.

[24] S. Wang, S. Kodagoda, S. and R. Ranasinghe, Road Terrain Type Classification based on Laser Measurement System Data, In: Australasian Conference on Robotics and Automation, (2012) Wellington, New Zealand, 1-6.

[25] J.G. Howe, D.C. Lee, J.P. Christos et al., Further Analysis of Potential Road/Terrain Characterization Rating Metrics, Journal of Commercial Vehicles 114(2) (2005) 228-241.

[26] G. Reina, A. Milella and R. Galati, Terrain assessment for precision agriculture using vehicle dynamic modelling, Biosystems Engineering 162 (2017) 124-139.

[27] A. Milella A., and G. Reina, 3D reconstruction and classification of natural environments by an autonomous vehicle using multi-baseline stereo, Intelligent Service Robotics, Vol. 7 (2), pp 79-92.

[28] J.C. Fernandez-Diaz, Characterization of surface roughness of bare agricultural soils using lidar, PhD, University of Florida, 2010

[29] G. Rill, Road Vehicle Dynamics Fundamentals and Modeling, CRC Press Taylor & Francis Group, 2011

[30] SEMI Draft Document 4622 (2008), Guide for estimating the Power Spectral Density Function and related finish parameter from surface profile data. Revision of SEMI MF1811-0704

[31] J.G. Proakis and D.G. Manolakis, Digital Signal Processing: Principles, Algorithms and Applications, Prentice-Hall International Inc., 3rd edition, 1996

[32] A. Brandt, Noise and Vibration Analysis, John Wiley & Sons, 2010

**Appendix**

**Proof** of Eq. (8)

Given a *non*-linear function $Y = g(X)$ of random variable $X$ and placing $\mu_X = E[X]$, by expanding in Taylor series $g(X)$ around $\mu_X$, one can obtain:

$$g(X) \approx g(\mu_X) + \frac{\partial}{\partial X}g(X)|_{\mu_X}(x - \mu_X) + \frac{\partial^2}{\partial^2 X}g(X)|_{\mu_X}(x - \mu_X)^2 + higher\ order\ terms$$

Thus, truncating the series to the first order, one can get:

$$E[g(X)] = \int g(X)f_X(x)dx = \int g(\mu_X)f_X(x)dx + \int \frac{\partial}{\partial X}g(X)|_{\mu_X}(x - \mu_X)f_X(x)dx = g(\mu_X)$$

Remembering that $\int f_X(x)dx = 1$, and moreover:

$$\int \frac{\partial}{\partial X}g(X)|_{\mu_X}(x - \mu_X)f_X(x)dx = \frac{\partial}{\partial X}g(X)|_{\mu_X}\int x f_X(x)dx - \frac{\partial}{\partial X}g(X)|_{\mu_X}\mu_X \int f_X(x)dx = 0$$

then, the first row of Eq. (8):

$$E[g(X)] = g(\mu_X)$$

To proof the second equation, one can start by the mean square value of $g(X)$:

$$E[g^2(X)] = \int g^2(X)f_X(x)dx$$

$$= \int g^2(\mu_X)f_X(x)dx + 2\int g(\mu_X)\frac{\partial}{\partial X}g(X)|_{\mu_X}(x - \mu_X)f_X(x)dx$$

$$+ \int \left[\frac{\partial}{\partial X}g(X)|_{\mu_X}\right]^2 (x - \mu_X)^2 f_X(x)dx$$

from which

$$E[g^2(X)] = g^2(\mu_X) + \left[\frac{\partial}{\partial X}g(X)|_{\mu_X}\right]^2 \sigma_X^2$$

and then the variance is

$$Var[g(X)] = \left[\frac{\partial}{\partial X}g(X)|_{\mu_X}\right]^2 \sigma_X^2$$